\documentclass[aps,prb,twocolumn,showpacs,groupedaddress]{revtex4-1}

\usepackage{graphicx}
\usepackage{bm}

\begin{document}


\title{Full counting statistics for SU($N$) impurity Anderson model}


\author{Rui Sakano$^1$, Akira Oguri$^2$, Takeo Kato$^3$ and Seigo Tarucha$^1$}
\affiliation{$^1$Department of Applied Physics, University of Tokyo, Bunkyo, Tokyo, Japan \\
$^2$Department of Physics, Osaka City University, Sumiyoshi, Osaka, Japan \\
$^3$Institute for Solid State Physics, University of Tokyo, Kashiwa, Chiba, Japan}



\begin{abstract}
We analyze the full counting statistics of a
multiorbital Kondo effect in a quantum dot
with the SU($N$) symmetry in the framework of the renormalized perturbation theory.
The current probability distribution function is calculated for an arbitrary 
dot-site Coulomb repulsion $U$ in the particle-hole symmetric case.
The resulting cumulant up to the leading nonlinear term of applied bias voltages
indicates two types of electron transfer,
respectively carrying 
charge $e$ and $2e$, with different $N$-dependences.
The cross correlation between different orbital currents shows exponential
enhancement with respect to $U$, which directly addresses formation of
the orbital-singlet state.
\end{abstract}


\maketitle


Experimental realization of the Kondo state in mesoscopic
devices has promoted further research of the Kondo effect
since it enables one to
access many-body effects in out-of-equilibrium under
finite bias voltages.
\cite{PhysRevLett.81.5225,vanderWiel22092000}
Recent theoretical studies of nonequilibrium noise for Kondo
dots
\cite{PhysRevLett.97.086601,PhysRevB.73.233310,PhysRevLett.100.036603,PhysRevLett.100.036604,PhysRevB.80.233103,JPSJ.79.044714,PhysRevB.83.075440}
have predicted enhancement of shot noise with
a fractional Fano factor,
and have stimulated subsequent experimental studies.
\cite{PhysRevB.77.241303,NaturePhys5.208,PhysRevLett.106.176601}

Deeper understanding of the nonequilibrium Kondo effect is brought in
by the current distribution function, which includes
higher-order cumulants beyond the first and second cumulant
(average current and noise power).
It, however, is still challenging
to calculate the cumulant generating function (CGF)
in the nonequilibrium Kondo states.
Recently, Komnik, Gogolin and Schmidt have derived
current probability distribution
for the SU(2) Anderson impurity
\cite{PhysRevLett.97.016602,PhysRevB.75.235105,PhysRevB.73.195301,PhysRevB.76.241307} 
from the general formulation of the full counting statistics (FCS).
\cite{RevModPhys.81.1665,PROP:PROP200610305}
They have clarified that the nonequilibrium backscattering current
is composed of two types of electron transfer due to 
a single quasiparticle and a pair of quasiparticles,
carrying charge $e$ and $2e$ respectively.

In the present work, we extend the FCS approach to multiorbital Kondo dots,
which have been experimentally investigated
in vertical dots,\cite{PhysRevLett.93.017205}
carbon nanotubes,\cite{NaturePhys5.208}
and double dots.\cite{Wilhelm2000668}
We consider an SU($N$) impurity Anderson model
as a prototype model for examining multiorbital effects.
By employing the renormalized
perturbation theory (RPT)
\cite{PhysRevLett.70.4007,0953-8984-13-44-314},
we calculate 
the zero-temperature CGF for the entire strength of the dot-site Coulomb repulsion.
The RPT is based on a general idea of renormalization in quantum field theory,
and is consistent with several known results,
indicating its
validity up to terms with third order of applied bias voltage.
\cite{PhysRevB.64.153305,JPSJ.74.110}
Our calculation provides direct information
on an {\it orbital singlet}, i.e., a correlated electronic state involving different orbitals.

{\it Model}---
Let us consider a single quantum dot system
described by the SU($N$) impurity Anderson model
${\cal H}_A^{} = {\cal H}_0^{} +{\cal H}_T^{}+{\cal H}_U^{}$ with
\begin{eqnarray}
{\cal H}_{0}^{} &=& \sum_{k \alpha m} \varepsilon_{k \alpha m}^{} c_{k \alpha m}^{\dagger} c_{k \alpha m}^{}
 + \sum_m  \epsilon_{dm}^{} d_m^{\dagger} d_m^{} \;, \\
{\cal H}_T^{} &=& \sum_{k \alpha m} \left( v_{\alpha}^{} d_m^{\dagger} c_{k \alpha m}^{} + \mbox{H.c.} \right) \;, \\
{\cal H}_U^{} &=& \sum_{m<m'} U d_{m}^{\dagger}d_{m}^{} d_{m'}^{\dagger}d_{m'}^{} \;,
\end{eqnarray}
where $d_m^{}$ annihilates an electron in the dot level $\epsilon_{dm}^{}$ with
orbital $m=1,2,\cdots, N$,
$c_{k \alpha m}^{}$ annihilates a conduction electron with moment $k$
and orbital $m$ in lead $\alpha = L,R$,
and $U$ is the dot-site Coulomb repulsion.
Here, orbital includes spin, 
and thus $N$ is even.
The intrinsic level width of the dot levels owing to tunnel coupling $v_{\alpha}$,
is given by $\Gamma = \sum_{\alpha} \pi \rho_c |v_{\alpha}|^2$
and the density of state of the conduction electrons $\rho_c$.
For simplicity,
the symmetric lead-dot coupling $v_L^{} = v_R^{}$
and the particle-hole symmetry $\epsilon_{dm}^{} = -(N-1)U/2$
are assumed.
The chemical potentials $\mu_{L/R}^{} = \pm V/2$,
satisfying $\mu_L^{} - \mu_R^{} = V(\geq 0)$,
are measured relative to the Fermi level which is defined 
at zero voltage $V=0$.
We take a unit, $\hbar = k_B = e =1 $, throughout this paper.

{\it Full counting statistics}---
The probability distribution $P({\bm q})$ 
of the transferred charge ${\bm q} = (q_1^{}, q_2^{}, \cdots, q_N^{})$
with orbital subscript
across the dot during a time interval ${\cal T}$
provides current correlation function of all orders.
In this paper,
we define the transferred charge operator as
$\hat{q}_m^{} \equiv n_{Lm}^{} (-{\cal T}/2) - n_{Lm}^{} ({\cal T}/2)$,
where $n_{Lm}^{} (t)$ is the electron number in the left lead.
In order to discuss
the correlation functions systematically,
we calculate the CGF
$\ln \chi \left( {\bm \lambda} \right) = \ln \sum_{\bm q} e^{i {\bm \lambda} \cdot {\bm q}}  P({\bm q})$
in the Keldysh formalism,
\cite{PhysRevB.70.115305}
\begin{eqnarray}
\ln \chi \left( {\bm \lambda} \right)
= \ln \left\langle 
 T_{C} S_C^{\lambda}
\right \rangle ,
\label{eq:CGF}
\end{eqnarray}
where
$S_{C}^{\lambda} = T_{C} \exp \left\{ -i \int_{C} dt \left[ {\cal H}_T^{\lambda}(t) + {\cal H}_U^{}(t) \right] \right\}$
is the time evolution operator for an extended Hamiltonian 
${\cal H}_A^{\lambda} = {\cal H}_0^{} + {\cal H}_T^{\lambda}+{\cal H}_U^{}$,
$C$ is the Keldysh contour along
$[t: -{\cal T}/2 \to +{\cal T}/2 \to -{\cal T}/2]$,
$T_{C}$ is the contour ordering operator,
and ${\bm \lambda}=(\lambda_1^{},\lambda_2^{}, \cdots, \lambda_N^{})$
is the counting field.
Here, ${\cal H}_T^{\lambda}$ is given by
\begin{eqnarray}
{\cal H}_T^{\lambda} = \sum_{k m} \left[ v_L^{} e^{i\lambda_m^{}(t)/2} d_m^{\dagger} c_{k L m}^{} +
v_R^{} d_m^{\dagger} c_{k R m}^{} \right]
+ \mbox{H.c.} \;,
\end{eqnarray}
with the contour dependent counting-field 
defined by
$\lambda_m^{} (t)= \lambda_{m\mp}^{} \equiv \pm \lambda_m^{}$ 
for the forward and backward 
paths labeled,  
respectively, by ``$-$" and ``$+$".

In order to calculate the CGF in Eq.\ (\ref{eq:CGF}),
we make use of a procedure suggested by Komnik and Gogolin,
\cite{PhysRevLett.94.216601}
which is outlined below.
First, a more general function
$\chi ({\bm \lambda}_{-}^{},{\bm \lambda}_{+}^{} )$
with ${\bm \lambda}_{\mp}^{}=(\lambda_{1\mp}^{},\lambda_{2\mp}^{},\cdots,\lambda_{N\mp}^{})$
is introduced. 
It is basically
given by Eq.\ (\ref{eq:CGF})
but $\lambda_{m\mp}^{}$ is 
treated formally as an independent variable
assigned for each contour.
For the long time limit ${\cal T} \to \infty$ where 
the switching effects are negligible,
the general CGF is proportional to ${\cal T}$,
\begin{eqnarray}
\ln \chi \left( {\bm \lambda}_{-}^{},{\bm \lambda}_{+}^{} \right)
= -i {\cal T} \, {\cal U} \left( {\bm \lambda}_{-}^{},{\bm \lambda}_{+}^{} \right) ,
\label{eq:exCGF}
\end{eqnarray}
with the adiabatic potential
${\cal U} ( {\bm \lambda}_{-}^{}, {\bm \lambda}_{+}^{} )$.
Once the adiabatic potential is computed,
the statistics is recovered from
$\ln \chi ({\bm \lambda} )
= -i {\cal T} {\cal U} ( {\bm \lambda}, -{\bm \lambda} )$.
Performing the derivative of
Eqs.\ (\ref{eq:CGF}) and (\ref{eq:exCGF})
with respect to $\lambda_{m-}^{}$,
we obtain
\begin{eqnarray}
\frac{d}{d\lambda_{m-}} {\cal U} \left( {\bm \lambda}_{-}^{},{\bm \lambda}_{+}^{} \right)
= 
\lim_{{\cal T} \to \infty} 
\left\langle \frac{d}{d\lambda_{m-}} 
{\cal H}_T^{\lambda} \left(0_- \right) 
\right\rangle_{\lambda_{-}^{},\lambda_{+}^{}} , 
\label{eq:dAP}
\end{eqnarray}
where we use a notation
\begin{eqnarray}
\langle A(t) \rangle_{\lambda_{-}^{},\lambda_{+}^{}}^{}
=
\left\langle 
T_{C}^{}
S_C^{\lambda} A(t)
\right\rangle 
\big/ \chi ( {\bm \lambda}_{-}^{},{\bm \lambda}_{+}^{} ) .
\label{eq:exexpectation}
\end{eqnarray}
Equation (\ref{eq:exexpectation}) presents an expectation for Hamiltonian
${\cal H}_A^{\lambda}$ and the Wick's theorem is applicable.
The right hand side of Eq. (\ref{eq:dAP}) can be expressed 
in terms of the Green's function,
\begin{eqnarray}
&& \frac{d}{d\lambda_{m-}^{}} {\cal U} ({\bm \lambda}_{-},{\bm \lambda}_{+}) \nonumber \\
&& \quad = \frac{\left| v_{L}^{} \right|^2}{2} \sum_k \int \frac{d\omega}{2\pi}
\left[ e^{-i \bar{\lambda}_m^{}/2} G_{dm}^{\lambda -+} \left( \omega \right) g_{Lkm}^{0+-} \left( \omega \right) \right. \nonumber \\
&&\quad \qquad \qquad \qquad \qquad - \left. e^{i \bar{\lambda}_m^{}/2} g_{Lkm}^{0-+} \left( \omega \right) G_{dm}^{\lambda +-} \left( \omega \right) \right] , \label{eq:dAP2}
\end{eqnarray}
with $\bar{\lambda}_m^{} = \lambda_{m-}^{} - \lambda_{m+}^{}$.
$g_{k\alpha m}^{0-+}(\omega)= i2\pi \delta(\omega -\varepsilon_{k\alpha m}^{}) f_{\alpha}^{}(\omega)$ and
$g_{k\alpha m}^{0+-}(\omega)= -i2\pi \delta(\omega -\varepsilon_{k\alpha m}^{}) [1 - f_{\alpha}^{}(\omega)]$ are
the lesser and greater parts of the Green's function for
electrons in lead $\alpha$
with the Fermi distribution function
$f_{\alpha}(\omega) = [e^{(\omega - \mu_{\alpha})/T} + 1]^{-1}$,
respectively.
For the long time limit ${\cal T} \to \infty$,
the dot Green's function is defined as
$\left\{
{\bm G}_{dm}^{\lambda} (\omega) 
\right\}_{\nu\nu'}
= -i\int d(t-t') e^{i\omega (t-t')} \langle T_{C}^{} 
d_m^{}(t_{\nu}) d_m^{\dagger} (t'_{\nu'}) 
\rangle_{\lambda_-,\lambda_+}^{}$.
Here, $\nu$ and $\nu'$ are the labels for the two Keldysh contours.

{\it Renormalized perturbation theory}---
The three basic parameters that specify the impurity Anderson model ${\cal H}_A^{}$ are $\epsilon_{dm}$, $\Gamma$ and $U$.
The low-energy properties can be
characterized by 
the quasiparticles with
the renormalized dot-level
$\widetilde{\epsilon}_{dm} = z \left[ \epsilon_{dm} + \Sigma_{dm}^r (0) \right]$,
renormalized level width
$\widetilde{\Gamma} = z \Gamma$,
and the renormalized interaction
$\widetilde{U} = z^2 \Gamma_{mm'}^{(4)}(0,0; 0,0) \quad (m \neq m')$,
where $\Sigma_{dm}^r (\omega)$ is the self-energy of the retarded Green's function
for the dot state: 
$G_{dm}^r (\omega) = [ \omega - \epsilon_{dm}^{} + i \Gamma - \Sigma_{dm}^r (\omega) ]^{-1}$,
$z = [ 1 - \partial \Sigma_{dm}^{r}(\omega)/\partial \omega|_{\omega=0} ]^{-1}$ is the wave function renormalization factor,
and $\Gamma_{mm'}^{(4)}(\omega_1, \omega_2 ; \omega_3, \omega_4)$ is the local full four-point vertex function for the scattering of the electrons 
with the orbital $m$ and $m'$.
\cite{PhysRevLett.70.4007,0953-8984-13-44-314}
Note that these parameters are defined  
at equilibrium $V=0$ and ${\bm \lambda}=0$.
The replacement of the bare parameters 
with the renormalized ones
gives leading terms of the Hamiltonian
corresponding to the low-energy fixed point of the
Anderson model 
in Wilson's theory.\cite{PhysRevB.21.1003}
The perturbation theory in powers of $U$ can be reorganized as 
an expansion with respect to the renormalized interaction $\widetilde{U}$,
taking the free quasiparticle 
Green's function $\widetilde{g}_{dm}^r (\omega) = ( \omega - \widetilde{\epsilon}_{dm} + i \widetilde{\Gamma})^{-1}$ as the zero-order propagator.
In addition, three counter terms 
are introduced in order to prevent overcounting.
This procedure has enabled one to calculate the exact form of 
the Green's function in the absence of the counting fields 
$\left.{\bm G}_{dm}^{\lambda}(\omega)\right|_{{\bm \lambda}=0}$
at low energies up to terms of order 
$\omega^2$, $V^2$, and $T^2$.\cite{PhysRevB.64.153305,JPSJ.74.110}
There are explicit relations between the renormalized
parameters and the enhancement factor of susceptibilities
$z\widetilde{\chi}_{d}^{} = 1+ \widetilde{U} \widetilde{\rho}_{dm}^{}(0) ,
z\widetilde{\chi}_{c,d}^{} = 1- (N-1) \widetilde{U} 
\widetilde{\rho}_{dm}^{}(0)$,
and the Friedel's sum rule
$\pi n_{dm}^{} = \cot^{-1} \left( \widetilde{\epsilon}_{dm}^{}/ \widetilde{\Gamma} \right)$
with the electron occupation in orbital $m$ of the dot
$n_{dm}^{}=\langle d_m^{\dagger} d_m^{} \rangle$,
and the renormalized density of state
$\widetilde{\rho}_{dm}(\omega)= (\widetilde{\Gamma}/\pi)/[(\omega - \widetilde{\epsilon}_{dm}^{})^2+ \widetilde{\Gamma}^2]$.
In particular, for the particle-hole symmetric case ($n_{dm}^{}=1/2$),
the renormalized parameters can be expressed in a simple form as,
$\widetilde{U}/( \pi \widetilde{\Gamma} ) = R_N^{} -1$ and $\widetilde{\epsilon}_{dm}=0$
with the Wilson ratio $R_N^{} = N/[(N-1)+{\widetilde{\chi}}_{c,d}^{}/{\widetilde{\chi}}_d^{}]$.
$\widetilde{\Gamma}$ can be considered as the Kondo temperature as,
$T_K = \pi \widetilde{\Gamma}/4$.
In this paper,
we evaluate
the renormalized parameters
by the Bethe ansatz exact solution (BAE)
\cite{Wiegmann1980163,Kawakami1981483}
and
the numerical renormalization group (NRG) calculation
\cite{PhysRevB.21.1003,PhysRevB.82.115123}.

Let us now apply the RPT to calculation of
the dot Green's function ${\bm G}_{dm}^{\lambda}(\omega)$
for the extended Hamiltonian ${\cal H}_A^{\lambda}$.
The dot Green's function is given by
\begin{eqnarray}
{\bm G}_{dm}^{\lambda} (\omega)
= z \widetilde{{\bm G}}_{dm}^{\lambda} (\omega)
= z \left[ { \widetilde{{\bm g}}_{dm}^{\lambda} (\omega)}^{-1}
- \widetilde{{\bm \Sigma}}_{dm}^{\lambda} (\omega) \right]^{-1} \label{eq:EXGF},
\end{eqnarray}
where the zero-order part is given by
\cite{PhysRevB.73.195301}
\begin{eqnarray}
\widetilde{g}_{dm}^{\lambda --} (\omega) &=&
\left[ \omega + i\widetilde{\Gamma} \left( f_L - 1/2 \right) + i\widetilde{\Gamma} \left( f_R - 1/2 \right) \right] \Big/ {\cal D}_m^{}, \nonumber \\
\widetilde{g}_{dm}^{\lambda -+} (\omega) &=&
\left[ i e^{i \bar{\lambda}_{m}^{}/2} \widetilde{\Gamma} f_L + i \widetilde{\Gamma} f_R \right] \Big/{\cal D}_m^{} , \nonumber \\
\widetilde{g}_{dm}^{\lambda +-} (\omega) &=&
- \left[ i e^{i \bar{\lambda}_m^{}/2} \widetilde{\Gamma} \left( 1-f_L \right) + i \widetilde{\Gamma} \left( 1 - f_R^{} \right) \right] \Big/ {\cal D}_m^{} , \nonumber \\
\widetilde{g}_{dm}^{\lambda ++} (\omega) &=&
\left[ - \omega + i\widetilde{\Gamma} \left( f_L - 1/2 \right) + i\widetilde{\Gamma} \left( f_R - 1/2 \right) \right] \Big/ {\cal D}_m^{}, \nonumber \\
\label{eq:EX0GF}
\end{eqnarray}
with
\begin{eqnarray}
{\cal D}_m^{} \left( \omega \right) = \omega^2 + \widetilde{\Gamma}^2
+ \widetilde{\Gamma}^2 \left[ \left( e^{- i \bar{\lambda}_m^{}/2 } - 1 \right) \left( 1 - f_L^{} \right) f_R^{} \right. \nonumber \\
+ \left. \left( e^{i \bar{\lambda}_m^{}/2} - 1 \right) 
\left( 1 - f_R^{} \right) f_L^{} \right] .
\end{eqnarray}
The remainder part of the {\it renormalized} self-energy
can be readily calculated in the second order perturbation in $\widetilde{U}$,
at $T=0$ up to $\omega^2, \omega V$ and $V^2$, as
\begin{eqnarray}
{\bm \Sigma}_{dm}^{\lambda} (\omega) =
\frac{-i}{8 \widetilde{\Gamma}} \left(\frac{\widetilde{U}}{\pi \widetilde{\Gamma}} \right)^2
\left[
\begin{array}{cc}
A_m(\omega, V) & B_m^{}(\omega, V) \\
-B_m^{\ast}(-\omega, -V) & A_m(\omega, V)
\end{array}
\right],
\label{eq:EXSE}
\end{eqnarray}
with
\begin{eqnarray}
A_m (\omega,V) &=&
 (N-1) \left[ a\left(\omega, 3V/2 \right) + 3\, a\left( \omega, V/2 \right) \right.  \nonumber \\
 && \quad+ \left. 3\, a\left( \omega, -V/2 \right) + a\left( \omega,-3V/2 \right) \right], \\
B_m(\omega, V)
&=& \sum_{m' (\neq m)}
\left\{ e^{-i \bar{\lambda}_{m'}^{}/2} b\left( \omega,3V/2 \right) \right. \nonumber \\
&& \quad + \left[ 2 + e^{i\left( \bar{\lambda}_m - \bar{\lambda}_{m'} \right)/2} \right] b \left( \omega, V/2 \right) \nonumber \\
&& \quad + \left[ 2 e^{i \bar{\lambda}_{m}/2} + e^{i \bar{\lambda}_{m'} /2} \right] b\left( \omega, - V/2 \right) \nonumber \\
&& \qquad \quad  + \left. e^{i ( \bar{\lambda}_{m} + \bar{\lambda}_{m'})/2} b \left( \omega,-3V/2 \right) \right\} .
\end{eqnarray}
Here,
$a(\omega,x)=- \frac{1}{2}(\omega - x)^2 \mbox{sgn}(-\omega + x)$, and,
$b(\omega,x)=(\omega -x)^2 \theta(-\omega +x)$.
For ${\bm \lambda} = 0$, the exact expression of usual self-energy
up to $\omega^2, \omega V$ and $V^2$ is reproduced.
\cite{PhysRevB.64.153305,JPSJ.74.110,PhysRevB.83.075440}
Substituting 
Eqs.\ (\ref{eq:EX0GF}) and (\ref{eq:EXSE}) into Eq.\ (\ref{eq:EXGF}), 
we readily obtain an expression for the Green's function
up to $\omega^2, \omega V$ and $V^2$.

{\it Results and discussion}---
Substituting the obtained Green's function into Eq.\ (\ref{eq:dAP2})
and integrating them respect with $\lambda_{m-}^{}$,
the CGF 
is derived at $T=0$
up to $V^3$, as $\ln \chi ({\bm \lambda})
= {\cal F}_0^{} + {\cal F}_1^{} + {\cal F}_2^{}$
with
\begin{eqnarray}
{\cal F}_0
&=& \frac{\cal T}{2\pi} \sum_m \int_{-V/2}^{V/2} d\omega
\ln\left[ 1 + \frac{\widetilde{\Gamma}^2}{\omega^2 + \widetilde{\Gamma}^2 } \left( e^{i \lambda_m^{} } -1 \right) \right] , \\
{\cal F}_1
&=& \frac{(N-1){\cal T}}{24 \pi} \frac{V^3}{\widetilde{\Gamma}^2} \left( R_N^{} -1 \right)^2 \sum_m \left( e^{-i\lambda_m} -1 \right) , \\
{\cal F}_2
&=& \frac{{\cal T}}{6 \pi}  \frac{V^3}{\widetilde{\Gamma}^2} \left( R_N^{} -1 \right)^2 \sum_{(m \neq m')} \left[ e^{-i( \lambda_m^{} + \lambda_{m'}^{}) } -1 \right] .
\end{eqnarray}
Here, $\sum_{(m \neq m')}$ takes sum of all combination of $m$ and $m'$ without $m=m'$.
${\cal F}_0^{}$ is the CGF of the zero-order part of the RPT,
${\cal F}_1^{}$ is the {\it single}-quasiparticle backscattering process
carrying charge $e$,
and ${\cal F}_2^{}$ is the {\it two}-quasiparticle backscattering process
where the two quasiparticles with different orbitals make a singlet state,
carrying charge $2e$. 
In particular, 
${\cal F}_1^{}$ and ${\cal F}_2^{}$ 
represent the reflection of quasiparticles 
by the residual interaction $\widetilde{U}$. 
This CGF for arbitrary $U$
corresponds to the SU($N$) extension of the hypothesis
presented by Gogolin, Komnik and Schmidt.
\cite{PhysRevLett.97.016602,PhysRevB.73.195301,PhysRevB.76.241307}

We now consider
the cumulant for full current
${\cal C}_n^{} = (-i)^n\frac{d_{}^n}{d\lambda_{}^n}\ln \chi(\lambda)$,
which is derived from the CGF with $\lambda_m^{}=\lambda$ for all $m$,
as
\begin{eqnarray}
{\cal C}_n^{}
= {\cal T} \left[ I_u \delta_{1n} + (-1)^n  (P_{b0} + P_{b1} + 2^{n} P_{b2}) \right], \label{eq:FullCCumulant}
\end{eqnarray}
where $I_u = NV/(2\pi)$ is the linear-response current 
and $\delta_{nn'}$ is the Kronecker's delta.
We can obtain
$P_{b0} = \frac{N}{24\pi}\frac{V^3}{\widetilde{\Gamma}^2}$
and
$P_{b1} =\frac{N(N-1)(R_N^{} -1)^2}{24\pi}\frac{V^3}{\widetilde{\Gamma}^2}$
from ${\cal F}_0^{}$ and ${\cal F}_1^{}$,
respectively.
These express the probability of the single-quasiparticle 
backscattering processes carrying charge $e$, per time.
Similarly,
that of the two-quasiparticle backscattering process
$P_{b2} = 
\frac{N(N-1)(R_N^{} -1)^2}{12\pi} \frac{V^3}{\widetilde{\Gamma}^2}$
is obtained from ${\cal F}_2^{}$.
In this process, the two quasiparticles with
the different orbitals 
form a singlet pair and carry charge $2e$, 
which causes the factor $2^n$ in Eq.\ (\ref{eq:FullCCumulant}).

\begin{figure}[tb]
\includegraphics[width=8.5cm]{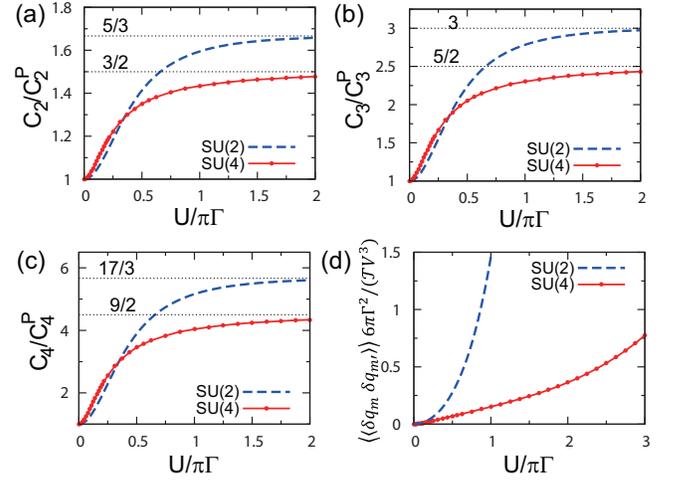}
\caption{
\label{fig:ratio-C}
(Color online)
(a) The FFIR ${\cal C}_n^{}/{\cal C}_n^P$ for $n=2$, (b) $n=3$, and (c) $n=4$,
as a function of the Coulomb repulsion $U$.
(d) The cross cumulant
between fluctuation of transmitted charge
with orbital $m$ and $m'(\neq m)$ 
as a function of the Coulomb repulsion $U$.
The blue solid line and the red broken line present
the SU(2) ($N=2$) case 
and the SU(4) ($N=4$) case, respectively.
}
\end{figure}

In order to extract universal properties
from the cumulant (\ref{eq:FullCCumulant}),
we consider the Fano-factor {\it inspired} ratio (FFIR)
\cite{PhysRevLett.97.016602}
for $n \geq 2$ and an arbitrary $U$, normalized in the form
\begin{eqnarray}
\frac{{\cal C}_n^{}}{{\cal C}_n^P}
= \frac{1+(1+2^{n+1})(N-1)(R_N^{} -1)^2}{1 + 5(N-1)(R_N^{} -1)^2}.
\label{eq:FIR}
\end{eqnarray}
Here, 
 ${\cal C}_{n}^P=(-1)^n I_b^{} {\cal T}$ ($n\geq 2$)
is the Poisson value of the CGF and
$I_b \equiv I_u - {\cal C}_1^{}/{\cal T}= P_{b0}+ P_{b1}+ 2 P_{b2}$ 
is the backscattering current.
Thus, the FFIR can be interpreted as
an average of the cumulants for the three backscattering current in this case.
Remarkably,
the FFIR is determined 
only by the two parameters, the Wilson ratio $R_N^{}$ and degeneracy $N$.
The Fano factor which corresponds to the FFIR for $n=2$, 
agrees with the previous works.
\cite{PhysRevB.80.233103,JPSJ.79.044714,PhysRevB.83.075440}
Evaluating $R_N^{}$ with the BAE for SU(2)($N=2$),
\cite{Wiegmann1980163,Kawakami1981483}
and with the NRG for SU(4)($N=4$),
\cite{PhysRevB.21.1003,PhysRevB.82.115123}
we plot the FFIR for noise ($n=2$), skewness ($n=3$)
and sharpness ($n=4$) as a function
of the Coulomb repulsion $U$
in Figs.\ \ref{fig:ratio-C}(a)-\ref{fig:ratio-C}(c), respectively.
In these figures,
with increase of $U$,
the FFIR crossovers from the Poisson value to the universal value of 
the SU(2) and SU(4) strong-coupling limit.
Note that the $n$ dependence, 
which enters through the factor $2^n$ in Eq.\ (\ref{eq:FullCCumulant}),
is caused by the two-quasiparticle process.
It makes the FCS of the Kondo systems 
quite different
from that of the noninteracting system $U=0$.
In the weak coupling limit $U \to 0$,
the Wilson ratio takes the value of $R_N^{}=1$,  
and thus the FFIR goes to ${\cal C}_n^{}/{\cal C}_n^P \to 1$.
In the strong-coupling limit $U \to \infty$,
the Wilson ratio approaches to
the value $R_N^{} \to N/(N-1)$,
and the FFIR takes a universal form
\begin{eqnarray}
{\cal C}_n^{} \big/ {\cal C}_n^P \to (N + 2^{n+1}) \big/ (N + 4) .
\label{eq:FFIRinKondo}
\end{eqnarray}
The explicit values of the FFIR
for several $N$ are given in the TABLE \ref{tab:uRatio}.
\begin{table}[tb]
\caption{\label{tab:uRatio}
The FFIR for $n=2,3$ and $4$
in the strong-coupling limit $U\to \infty$, given in Eq.\ (\ref{eq:FFIRinKondo}),
for several choices of degeneracy $N$.}
\begin{ruledtabular}
\begin{tabular}{cccccc}
$N$ & 2 & 4 & 6 & 8 & $\to \infty$ \\
\hline
${\cal C}_2^{}/{\cal C}_2^P$ & 5/3 & 3/2 & 7/5 & 4/3 & $\to 1$ \\
${\cal C}_3^{}/{\cal C}_3^P$ & 3 & 5/2 & 11/5 & 2 & $\to 1$ \\
${\cal C}_4^{}/{\cal C}_4^P$ & 17/3 & 9/2 & 19/5 & 10/3 & $\to 1$ \\
\end{tabular}
\end{ruledtabular}
\end{table}
For $N=2$, 
Eq.\ (\ref{eq:FFIRinKondo}) agrees with the result given 
by Gogolin and Komnik.\cite{PhysRevLett.97.016602}
In the limit of large degeneracy $N \to \infty$, however,
the FFIR approaches to the Poisson value ${\cal C}_n^{}/{\cal C}_n^P \to 1$,
even though the Coulomb repulsion has been taken first to be $U \to \infty$.
This is 
because the renormalization becomes weaker 
for larger $N$
and the two-quasiparticle process is suppressed $P_{b2} \to 0$.

We next consider cross cumulant
which is observed as cross correlation between different orbital currents,
and it may
enable one to directly observe
the two-quasiparticle scattering in experiments.
The generic form of the
cross cumulant between orbital currents is derived 
from the CGF
up to terms of $V^3$ as
\begin{eqnarray}
\langle \langle \delta q_m^k \delta q_{m'}^l \rangle \rangle
&=& (-i)^{k+l} \frac{\partial^{k+l} }{\partial \lambda_m^k \partial \lambda_{m'}^l}
\ln \chi \left( {\bm \lambda} \right) \nonumber \\
&=& (-1)^{k+l} \frac{{\cal T}}{6 \pi} \frac{V^3}{\widetilde{\Gamma}^2} \left( R_N^{} -1 \right)^2 \;,
\end{eqnarray}
for $k, l\geq 1$ and $m \neq m'$.
The cross cumulant for $(k,l)=(1,1)$
is plotted as a function of the Coulomb repulsion $U$
in Fig.\ \ref{fig:ratio-C} (d).
In the strong-coupling region,
the cross cumulant is inversely proportional
to the square of the renormalized level width as 
$\langle \langle \delta q_m^k \delta q_{m'}^l \rangle \rangle \propto 1/\widetilde{\Gamma}^2$,
and, thus,
increases exponentially with increase of the Coulomb repulsion 
as shown in Fig.\ \ref{fig:ratio-C} (d).
The positive cross cumulant is a signature of
orbital-singlet states traveling through the dot.
Larger orbital degeneracy makes renormalization weaker
and suppresses the two-quasiparticle scattering.
Therefore, the cross cumulant
becomes the largest 
in the SU($2$) case and the orbital degeneracy suppresses the correlation.
We note that
there is no higher order cross cumulant
such as 
$
\langle \langle \delta q_m^j \delta q_{m'}^k \delta q_{m''}^l\rangle \rangle
$ for $m\neq m' \neq m''$ 
 at low bias voltages determined by the terms up to order $V^3$ 
in the  particle-hole symmetric case 
even in the presence of the orbital degeneracy $N>2$.
Naively, it seems that
the cross correlation can be observed
in double dots with interdot Coulomb repulsion,\cite{Wilhelm2000668}
or dots connected to ferromagnetic leads with opposite polarizations.

Finally, 
we comment on cumulants of the current for orbital $m$,
${\cal C}_n^m = (-i)^n \frac{d_{}^n}{d\lambda_m^n} 
\ln \chi ({\bm \lambda}) = (-1)^n I_b^m$
with the backscattering current
$I_b^m=V/h - {\cal C}_1^m$.
The cumulant ${\cal C}_n^m$ always takes the Poisson value 
${\cal C}_n^m/{\cal C}_n^{mP}=1$ 
in the particle-hole symmetric case 
for the contribution up to $V^3$
because there is no scattering of two quasiparticles
with the same orbital in this case.

{\it Summary}---
We have investigated the FCS
of a multiorbital Kondo dot described by
the particle-hole symmetric SU($N$) impurity Anderson model.
Using the RPT,
we derived the CGF for arbitrary Coulomb repulsion up 
to terms of order $V^3$.
The dot-site Coulomb repulsion
induces quasiparticle's orbital-singlet pairs 
carrying charge $2e$ in the backscattering current.
This process characterizes 
quantum fluctuations of the current in the correlated dot 
and is particularly manifest in
the cross correlation between orbital current.
It is also found that there is
no electron entangled state
carrying more than three quasiparticles
in the particle-hole symmetric case in current up to order $V^3$,
even in the presence of large orbital degeneracy $N>2$.

The authors thank Y. Utsumi, K. Kobayashi, A. C. Hewson,
A. O. Gogolin, Y. Okazaki, R. S. Deacon,
S. Iwabuchi and T. Fujii
for fruitful discussion.
This work was supported by
the JSPS through its FIRST program,
the JSPS Grant-in-Aid
for JSPS Fellows
and Scientific Research C (No. 23540375) and S (No. 19104007),
and, the Grant-in-Aid for Young Scientists B
(No. 21740220)
from MEXT, Japan.
Numerical computation was partly carried out
at Yukawa Institute Computer Facility.


%

\end{document}